\documentclass[conference]{IEEEtran}
\ifCLASSINFOpdf
\else
\fi
\usepackage{amsmath}
\usepackage{tikz}
\usepackage{amsmath}
\usepackage{dblfloatfix}
\usepackage{lipsum}
\usepackage{graphicx}
\usepackage{url}
\usepackage[breaklinks]{hyperref}
\usepackage{multirow}
\usepackage{pbox}
\usepackage{tabularx}

\begin{document}
%
\title{Beyond Trolling: Malware-Induced Misperception Attacks on Polarized Facebook Discourse}

\author{\IEEEauthorblockN{Filipo Sharevski}
\IEEEauthorblockA{DePaul University\\ fsharevs@cdm.depaul.edu}
\and
\IEEEauthorblockN{Paige Treebridge}
\IEEEauthorblockA{DePaul University\\ 
ptreebri@cdm.depaul.edu}
\and
\IEEEauthorblockN{Peter Jachim, Audrey Li, Adam Babin}
\IEEEauthorblockA{Divergent Design Lab\\ 
www.divergentdesignlab.org}
\and
\IEEEauthorblockN{Jessica Westbrook}
\IEEEauthorblockA{DePaul University\\ 
jwestbroo@cdm.depaul.edu}}


%


\IEEEoverridecommandlockouts

\maketitle

\begin{abstract}
Social media trolling is a powerful tactic to manipulate public opinion on issues with a high moral component. Troll farms, as evidenced in the past, created \textit{fabricated} content to provoke or silence people to share their opinion on social media during the US presidential election in 2016. In this paper, we introduce an alternate way of provoking or silencing social media discourse by manipulating how users perceive \textit{authentic} content. This manipulation is performed by man-in-the-middle malware that covertly rearranges the linguistic content of an authentic social media post and comments. We call this attack Malware-Induced Misperception (MIM) because the goal is to socially engineer \textit{spiral-of-silence} conditions on social media by inducing perception. We conducted experimental tests in controlled settings ($N=311$) where a malware covertly altered selected words in a Facebook post about the freedom of political expression on college campuses. The empirical results (1) confirm the previous findings about the presence of the \textit{spiral-of-silence} effect on social media; and (2) demonstrate that inducing misperception is an effective tactic to silence or provoke targeted users on Facebook to express their opinion on a polarizing political issue.  
\end{abstract}


%

\section{Introduction}
Social medial trolling became widely known phenomenon, reaching organized dimension in 28 countries by 2017 \cite{DiResta}. Trolling refers to users who respond to social media posts with fabricated and often inflammatory posts and comments to get a rise out of users \cite{Hanson}. Organized trolling campaigns usually target specific populations, attempting to sway the opinion of entire groups, for example, domestic voters. Over time, the coordinated  "nudging" of public opinion has become systematized, from military units that experiment with psychological operations to strategic communication firms that take contracts from governments for social media campaigns aiming to induce misperception \cite{Bradshaw}.

A malicious actor interested in trolling seeks to "nudge" opinions on polarizing or controversial issues discussed via social media, e.g. election campaigns, climate change, vaccination, immigration policy, reproductive health and freedom of political expression. An overt strategy is to do what the Russian trolling army did in 2016: manufacture a large number of political trolling posts to rile up Americans \cite{Thompson}, \cite{DiResta}. This is an arduous task as it requires many people and resources in order to be successful (i.e. a bot network and a lot of fabricated content) \cite{Paavola}. There is also a risk that the social media administrators will remove any suspicious posts \cite{Alba}. The malicious actors will likely continue to search for covert alternatives to manipulate public opinion through social media in a more targeted fashion. One alternative is to still use fabricated content and induce "information gerrymandering" \cite{Stewart}. This still requires a large network of people who need to create and strategically infuse posts and comments on social media. A more economic alternative is a malware that acts as a man-in-the-middle in exchanging online information and manipulates how authentic content is \textit{perceived} by targeted individual. The advantage of the malware is that it is  platform-agnostic (i.e. can work on Facebook, Twitter, or Reddit) and can be strategically packaged as a web browser extension or a third-party social media application for smartphones.

Studies on manipulating online information point that \textit{induced misperceptions} represent an effort of a malicious actor to "lead an individual towards making false or implausible interpretations of a set of true facts" \cite{Benkler}. In the same manner, this malware covertly swaps, rearranges, or removes words presented to an individual to induce interpretation of a set of true facts to the objective of a malicious actor. Using a malware to induce misperception, to our knowledge, is a zero-day social engineering attack because it allows the targeted individual to verify the authenticity of online information thus bypassing all conventional cues people use to detect "phishy" or fabricated content \cite{Ferreira}. Like phishing, the malware also employs the psychological principles of persuasion to obtain individuals' assets (e.g. system permissions) but not to damage the local files or exfiltrate data \cite{Cialdini}. Instead, the goal is to use the system permissions to covertly manipulate textual data in transit and induce interpretation of legitimate content biased towards the objective of the malicious actor, e.g. poach disgruntled workers or bias voters \cite{Savvas}.  

This paper introduces the concept of malware-induced misperception and reports a test of the attack on polarized discourse on Facebook. The goal was to investigate whether this malware can be used to engineer or disrupt the \textit{spiral-of-silence} effect on social media, that is, to manipulate how users perceive an authentic Facebook post and comments instead of using any fake information or inflammatory content. The spiral-of-silence theory argues that individuals fear becoming socially isolated, and as a consequence, they constantly monitor the public opinion climate on mass media to determine whether the majority shares their own opinions or not \cite{Noelle-Neumann}. If the individuals \textit{perceive} that their own opinion is in the minority, they end up silencing themselves, especially when discussing polarizing or controversial issues. The theory, originally developed for face-to-face interpersonal communication, is also applicable in social media settings \cite{Matthes}. 

A sample of 311 participants was randomly assigned to a control and treatment group. The participants in the control group were exposed to a legitimate Facebook post and comments in a web browser while the participants in the treatment group saw a malware-manipulated version of the same Facebook post and comments. The discourse was on the polarizing issue of freedom of speech on college campuses \cite{Beauchamp}, \cite{Peters}. The malware was packaged as a web browser extension as a low-cost option that allowed controlled use only in laboratory settings (alternative packaging is also discussed in the paper) \cite{Newman}. The results show that the malware could successfully engineer the spiral-of-silence effect for individuals on the far ends of the political spectrum. The results are in line with the previous findings that people with divergent opinions "use Facebook as a forum to monitor the prevailing public opinion on important polarizing issues without expressing their own comments \cite{Gearhart}, \cite{Kwon}. In the reminder of the paper, Section 2 elaborates the social engineering background of the MIM attack. Section 3 discusses the spiral-of-silence theory underpinning the MIM attacker's social engineering strategy when applied to polarizing discourse on Facebook. Section 4 covers the study design and Section 5 presents the empirical results. Section 6 discusses the implications of materializing malware-induced misperceptions beyond social media and ways to counter these attacks. Section 7 concludes the paper. 

\section{Malware-Induced Misperception}
\subsection{Concept}
Conventional social engineering attacks target individuals' \textit{assets}, e.g. passwords or system privileges. These assets enable social engineers to obtain unauthorized access so as to damage or exfiltrate confidential data. For this purpose social engineers usually write various types of malware (e.g. adware, trojans, keyloggers, rootkits, etc.). The most common vector for malware delivery and installation is through "phishing", i.e. an email or a text where the social engineers employ various principles of persuasion to secretly obtain the target individual's compliance to run the malware code on their machine \cite{Ferreira}. The phishing campaigns can be massive and target the largest number of individuals possible or they can target specific and well-researched individual(s) \cite{Hardy}. Social engineering attacks are notoriously successful and abundant effort is invested in detecting suspicious content as well as training individuals to spot both massive and targeted or "spear" phishing emails \cite{Alsharnouby}, \cite{Khonji}. 

Because phishing attacks are low-cost/high-reward, social engineers have the possibility to try different persuasion routes and choose how to utilize the target individual's assets. In this paper we introduce a social engineering attack utilizing a malware that targets the integrity but not the confidentiality of the target individual's data. The attack is executed in two stages. First, like in conventional phishing, the target individual is persuaded to install a seemingly benign software plug-in, that is, yield their system privileges for manipulating textual data. Second, these privileges are used to covertly manipulate the linguistic content of the online communication the target individual exchanges through a browser or an email client. The goal of this covert manipulation, by contrast to conventional phishing, is to induce misperception about an event, news report, a communicating party, or a communication context ~\cite{Benkler}. Such an attack, to our knowledge, has not yet surfaced in the cyber realm. We therefore named it a Malware-Induced Misperception (MIM) attack. The covert linguistic manipulation of online communication is specific to a target individual (e.g. linguistic style, pragmatics, cultural norms, etc.), therefore, the MIM attack is more feasible in a spear phishing form. The attack is low-cost in that the malware could be packaged either as a browser extension, an email client "add-in" (e.g. Outlook), or perhaps in the future an entirely new application. The high-reward of the attack, if successful, is the opportunity to distort the target's mental picture or map of reality to establish psychological domination. 

Distorting individual's map of reality by inducing misperception has become a significant problem on social media over the past few years. Malicious actors like trolls, sock puppets, and alternative media flooded Facebook and Twitter prior to the US presidential election with rumors, fake news, and inflammatory comments with the objective to bias people and sway their votes \cite {Spangher}. After these efforts were shored by Facebook and Twitter, malicious actors proceeded with a strategic infusion of fabricated content for particular events and towards well-researched individuals in a tactic called "information gerrymandering" \cite{Stewart}. The idea is to manufacture echo chambers to create a (mis)perception that "most of the others were going to for the other party” (an improved version of Cambridge Analytica's strategy targeting voters in sway districts \cite{Granville}). In all of these cases, the malicious actors relied on a considerable number of people who fabricated these posts or relentlessly posted inflammatory comments on social media.

The MIM attack is inspired by these misperception campaigns but takes advantage of the social engineering tactics. The malware replaces the need for constantly fabricating content or infusing inflammatory social media posts and comments. The malware also elevates worries that the social media platform can detect a misperception campaign. Instead, the misperception takes place on a local machine or smartphone where the malware covertly rearranges the words and the "tone" of an authentic social media post while the targeted individual is reading it in real time. Studies on manipulating online information point that \textit{induced misperceptions} represent an effort of a malicious actor to "lead an individual towards making false or implausible interpretations of a set of true facts" \cite{Benkler}. By targeting authentic content, the malware allows the targeted individual to verify the facts and the credibility of a source thus bypassing all conventional cues people use to detect "phishy" content \cite{Ferreira}. The goal of the malware is to covertly manipulate the data in transit and induce interpretation of authentic content biased towards the objective of the malicious actor, e.g. bias voters, poach a high-profile target, or introduce fear, doubt, and uncertainty. 

\subsection{Implementation}
This malware can be packaged as a browser extension, an email client "add-in" (e.g. Outlook), or an entirely new application. The malware usually is disguised as seemingly benign (e.g an extension for accessibility support, Outlook add-in for managing email threads, or a lightweight, power-saving mobile app). This packaging/disguise is preferred because the malware requires text manipulation permissions that later will be leveraged for the MIM attack \cite{Vincent}. Developing extensions, add-ins, and apps is free and a benign software can pass all the security checks before publishing \cite{Newman}. For example, a browser extension variant of the malware can disguise the misperception-inducing logic and pass the security checks by posing as an "accessibility (a11y) extension" that claims the rewording is done to help non-native English speakers make sense of English slang on social media \cite{Jang}. An email add-in variant of the malware can pass the security checks, similarly, on the grounds of grouping and classifying social media email reports for better management through an Outlook client \cite{Tyler}. Certainly, the malware could be packaged as a third-party smartphone app that, for example, claims to reduce the battery usage by summarizing the content of social media posts \cite{Seals}.

The coordinated effort to sway people's opinions about polarizing issues on social media makes a compelling case for the MIM attack to be implemented either as a browser extension or a third-party social media app. For the purpose of our study we developed the malware as a browser extension in JavaScript as a more economic proof-of-concept variant. The goal was to investigate whether the malware can induce the spiral-of-silence effect on social media, that is, influence a target individual to divulge a comment or personal opinion on social media that they otherwise wouldn't post, fearing social isolation. We conducted a pilot study with 15 volunteer participants where we tested the malware's potential to induce misperception on a simple Facebook post. All participants were 18 years or older, regularly read and commented on Facebook posts through a web browser, and had prior knowledge of social engineering, phishing, and past social media trolling, misperception, and fake news campaigns. 

The preliminary question was to gauge whether participants are open to using browser extensions for standard utilities, for example an add-blocker or a sticky notes extension like "Stickies" \cite{Vincent}. Most of them responded they already do use various extensions that improve their productivity and install them almost immediately after downloading or start using a web browser on their computers. Some of the participants were aware that browser extensions could potentially contain spyware and affect their privacy or steal personal information like remembered passwords or credit cards, and they look for legitimate extensions only on the browser application stores. Some of them were aware of extensions that manipulate content, like the Facebook demetricator, that hides the number of likes on Facebook posts to enable a more immersive interaction with the social media platform \cite{Grosser}. None of them were aware of browser extensions that covertly rearrange text before it is rendered in a browser. This was an important feedback suggesting that it is plausible for a malicious actor to employ a legitimacy-by-design (seeming legitimate both in visual design and in what the user expects to see from a legitimate application) to persuade the target user to install a benign extension in the first place \cite{Newman}. 

The pilot participants first encountered an authentic Facebook post, shown in Figure 1, and reported that they are not inclined to comment on it, explaining that the post fits the campaign narrative of Mr. Sanders for the forthcoming US elections in 2020. The malware then was used to covertly swap the position of the words "Commander" and "Organizer," as shown in Figure 2, to induce misperception that Mr. Sanders is shifting his campaign strategy from peaceful to militaristic \cite{Scher}. Noticing that the accent of the post is on the "Commander in Chief" instead of "Organizer in Chief," the participants felt compelled to express concerns about Mr. Sanders' true intentions as a potentially future president and ask questions about this shift through comments. 

\begin{figure}[h]
\centering
  \includegraphics[width=0.8\columnwidth]{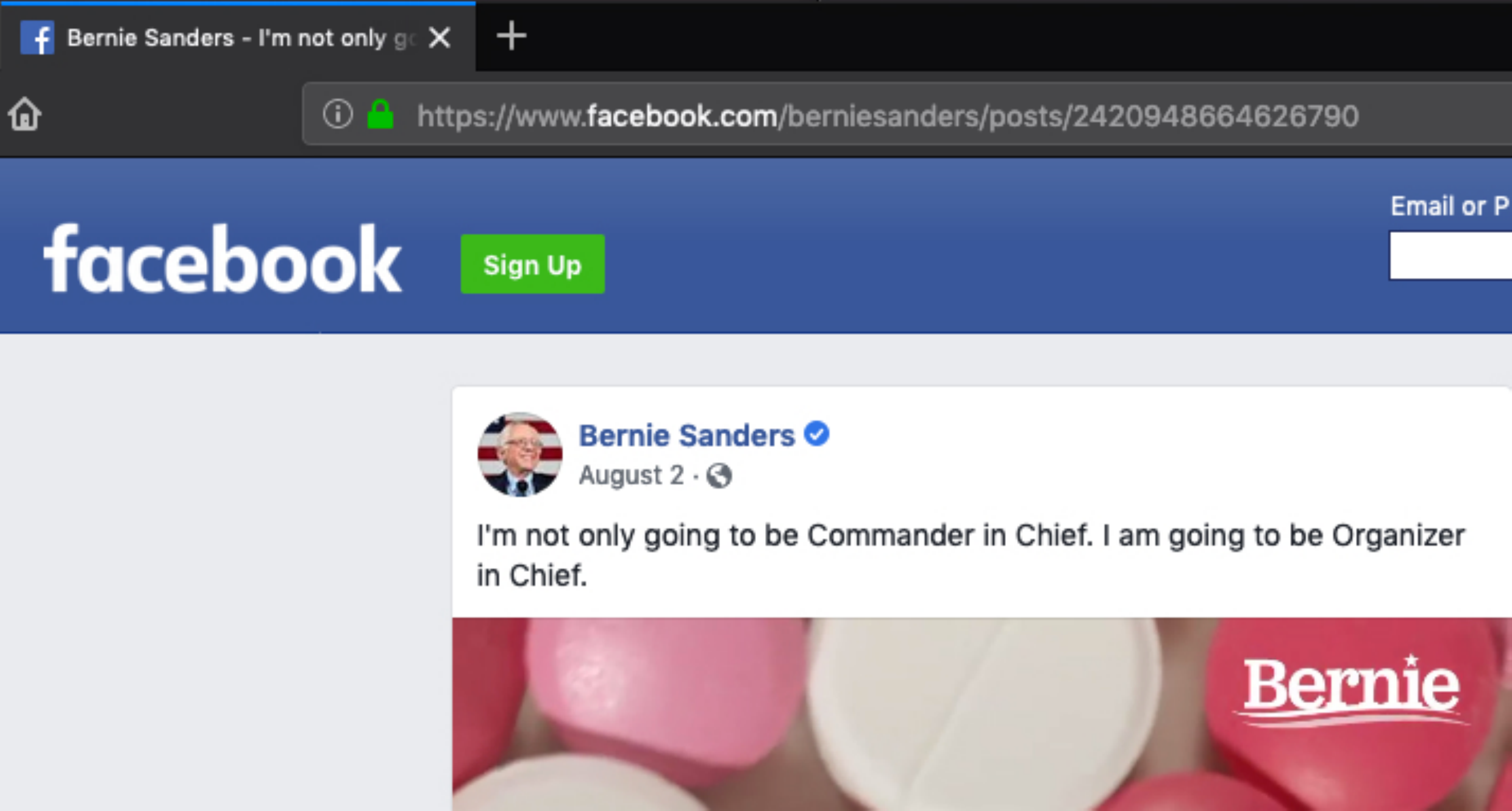}
  \caption{MIM extension "off"}
  \label{Fig2}
\end{figure}

\begin{figure}[h]
\centering
  \includegraphics[width=0.8\columnwidth]{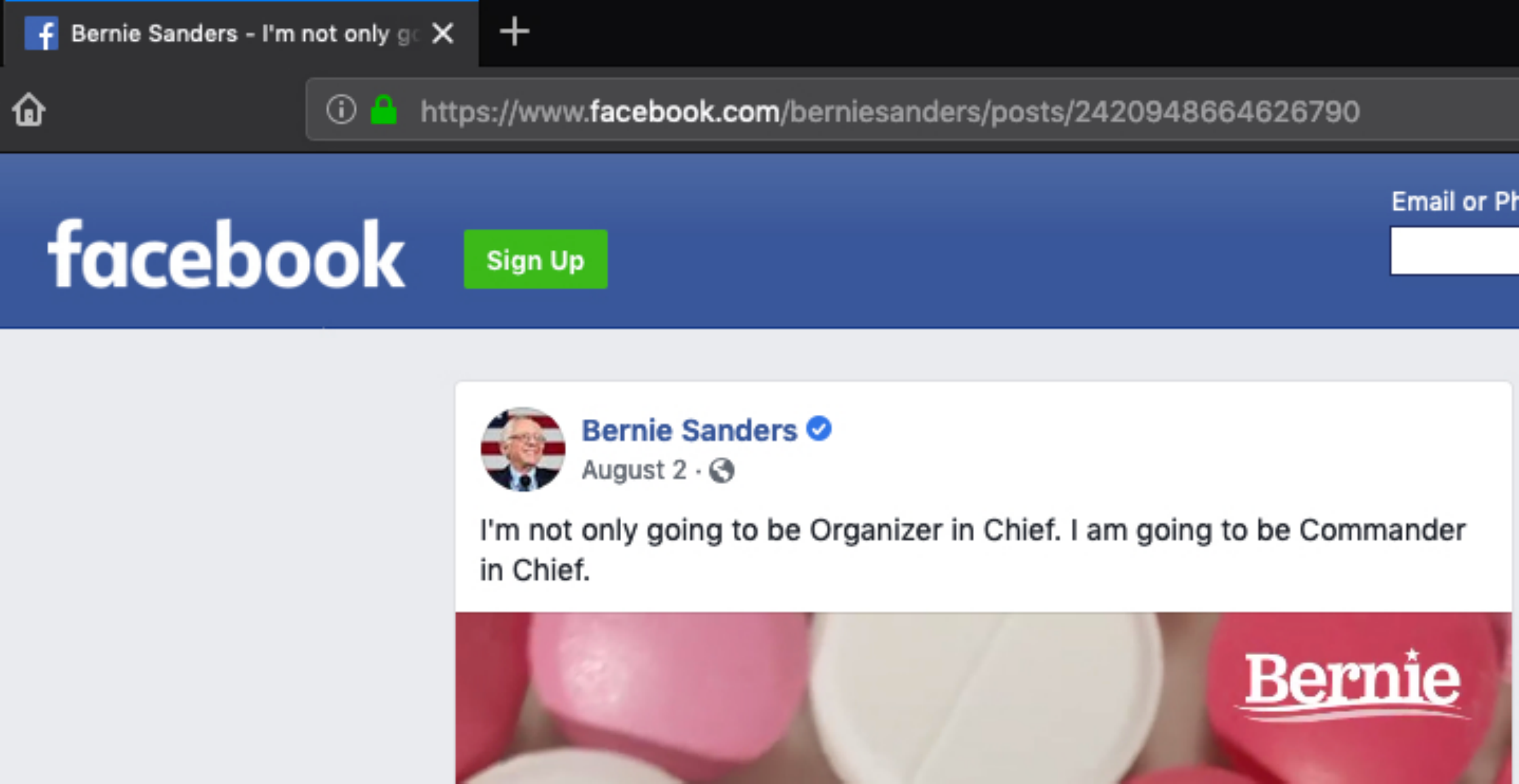}
  \caption{MIM extension "on"}
  \label{Fig3}
\end{figure}

An important feature of the malware is that it allowed the participants, trained in spotting "phishy" and inflammatory content, to verify the post (i.e. this is a valid campaign message by Senator Sanders) and verify the credibility of the sender (this is the official Facebook page of Senator Sanders \cite{Bernie}). The MIM attack in the pilot study successfully disrupted the spiral-of-silence effect by inducing misperception of the next steps of Senator Sanders's presidential campaign. This motivated us to test the potential of the MIM attack to socially engineer similar effects with a larger sample of participants on similar, but rather implicit, polarizing political discourse on Facebook.

\subsection{Threat Model} 
The MIM advantage, from the perspective of an attacker, is that the relationship between the target user and a web resource, or another person, can be manipulated without alerting any of the involved parties. The malware can be employed, for example, to influence a target user to divulge a comment or personal opinion on social media that they otherwise wouldn't post, fearing social isolation. MIM can be categorized as a threat where an adversarial group or nation-state (threat source) conducts externally-based electronic communication modification i.e. man-in-the-middle  attacks. MIM is a complex and micro-targeted attack and it requires a sophisticated level of expertise and well-resourced adversary \cite{NIST}. The intent for launching a MIM attack can be a low-intensity trolling campaign, provoking (or silencing) comments on social media for posts with a strong moral component. A target for MIM can be any public person like a political party leader, a celebrity, or a social media influencer, but MIM can equally target disgruntled employees, a spouse, or friends. The predisposing conditions for a successful MIM attack are: (1) a targeted user to install a software that can be dynamically modified to manipulate text (a web browser extension in our case); (2) the targeted user accesses social media regularly, Facebook in particular; and (3) the targeted user is interested in polarizing issues extensively discussed on social media. 

\subsection{Linguistic Manipulation Strategies}
The malware works on a string array of "valence words" and word replacement logic if a target word is detected on Facebook page. The malware parses the HTML document with a \texttt{findMatch()} function to detect a potential word match. If a match is detected, \texttt{findMatch()} returns the opposite valenced/target array word of the source word. A \texttt{textSwap()} function then replaces the occurrences of the initially detected word based on a configurable logic (all occurrences, only the first occurrence, or only if the occurrence is in the comments section of a Facebook page). This is the simplest, low-cost low-complexity version of the malware. A MIM attacker can implement more complex logic where the linguistic manipulation can take place only in certain parts of the Facebook content or only in Facebook posts reporting on a specific person or issue, for example, only campaign posts by Senator Sanders but not the other presidential candidates. The string array of "valence words" need not to be predefined in that an attacker could use natural language processing to analyze authentic Facebook content and adapt the linguistic rearrangement that makes the most sense in the context of target individuals' Facebook diet [redacted]. Using a Markov chain a model can be trained to choose replacement words based on an identified corpus of Facebook content. This natural language processing strategy was previously used by other researchers to generate a series of quotes that sound like President Trump’s State of the Union \cite{Downey}. 


\subsection{Social Media Vector}
The MIM attack differentiates itself from targeted ad campaigns on social media like the ones produced by Cambridge Analytica or requested by the UK Labour Party leaders to save on campaign costs and target only the party leader Jeremy Corbyn's Facebook account \cite{Baldwin}, \cite{Granville}. The attack is distinct from "information gerrymandering" where content is infused strategically in a social network to exploit the "homophily" - people's natural tendency to surround themselves with others who share their perspectives and opinions about the world (the "echo chambers" effect) \cite{Gillani}. While the aforementioned tactics aim to manipulate the perception of social media content, MIM doesn't require access to external user data nor uses ads or fabricated comments aimed to reinforce an echo chamber. Instead, MIM works directly on the social media post exploiting the main attention of a targeted user. This is beneficial to for micro-targeting individuals without worrying that the social media platform might detect the attack. 

As with the early period of political trolling, fake news, and alternative media, this creates a situation where people are left to resist and reject suspicious content by themselves. However, the malware could plausibly evade this detection because it preserves factual structure of the social media content. Even if someone is aware and carefully looking for inflammatory content or fake news the attack removes the grounds for such suspicion by working on authentic content \cite{Shao1}. In other words, the attack covertly "nudges" a targeted individual to make interpretations of a set of true facts to the objective of the malicious actor \cite{Benkler}. The MIM attack has the potential to be used for a purpose of trolling and spreading rumors, if the targeted words are aggression or produce misinformation. Nonetheless, the malware's primary goal of inducing misperception is the focus of the study.

\section{Spiral-of-Silence}
\subsection{Theoretical Background}
Spiral-of-silence theory, developed by Noelle-Neumann, argues that people use their media environment as a barometer for the prevailing climate of opinion on controversial issues \cite{Noelle-Neumann}. Printed newspapers and TV, and now the Internet and social media, operate as a social monitor by alerting the public about the perceived appropriateness of publicly expressing certain opinions. This is the case because society threatens with isolation those individuals who violate the societal consensus on values and goals. This consensus, expressed through the majority opinion in the media, influences how people form their individual opinion and action. Individuals whose opinions do not coincide with the majority opinion, as they perceive it, tend to silence their opinions, fearing social isolation \cite{Scheufle}. This silence effect results from one's perceptions of public opinion climates and susceptibility to social pressure. 

Numerous public opinion studies have applied spiral-of-silence theory to empirical examination \cite{Scheufle}, \cite{LinSalawen}, \cite{Matthes}. The primary dependent variable for the predisposing spiral-of-silence conditions in most of them is the \textit{willingness to express} one's opinion. As the original theory posits, human behavior, particularly the willingness to express one's opinion, is heavily directed by a fear of isolation that makes sanctions of denial of sympathy, and so forth, very powerful forms of influence. However, this significant variation in these predisposing conditions and thus the effects of the spiral-of-silence prompted a redefinition towards capturing one's \textit{willingness to self-censor}, defined as "the withholding of one's true opinion from an audience perceived to disagree with that opinion" \cite{Hayes}. In a social discourse, individuals can't simply stay silent, but instead look for a way to avoid expressing their opinion through some other methods. The self-censorship predisposing conditions are expressed through four opinion expression strategies individuals resort to when discussing issues with high moral component: (1) comment on the issue; (2) read or listen about the issue but choose not to comment; (3) ignore it; (4) tell someone else about it offline.

\subsection{Spiral-of-Silence on Social Media} 
The spiral-of-silence theory was developed for face-to-face communication and considers printed and televised mass media content. The Internet has changed the way people communicate and receive mass media - it provides anonymity and at the same time affords individuals access to diverse media content, autonomy, selectivity, and social media interactivity \cite{Gearhart2}. The fundamental change in interpersonal communication and media exposure prompted researchers to test the spiral-of-silence theory in the context of social media. Since social media interactions are anchored in real-world relationships, social media interactions are still vulnerable to fears of social isolation. Individuals online may express their opinions in ways that may "result in appearing unpopular or otherwise socially undesirable within the social media community" \cite{Metzger}. 

Our research builds upon existing studies that point to the validity of the spiral-of-silence effect on social media. A study examining how social media is used to express opinions on the issue of LGBT+ tolerance found that the spiral-of-silence phenomenon is present on Facebook \cite{Gearhart}. Testing how perceptions of surveillance contribute to an online spiral-of-silence in the wake of the Edward Snowden's revelations, authors in \cite{Stoycheff} found that the government's online surveillance programs may threaten the disclosure of minority views and contribute to the reinforcement of majority opinion. A study of discussion on nuclear power generation showed that the spiral-of-silence phenomenon exists on Twitter, too. Confirming the tenability of this theory in a social media context, a meta-analysis of the spiral-of-silence demonstrated that the relationship between opinion climate perception and opinion expression is as equally strong and robust on social media as it is in face-to-face communication \cite{Matthes}. 

\subsection{Spiral-of-Silence in Social Media on Political Issues}
People are regularly exposed to political content on social media. A Pew research report indicates that users on social media are more exposed to political perspectives dissimilar from their own than in face-to-face encounters \cite{Pew1}. Disagreements between users on social media on political topics is very common. For example, 73\% of the surveyed users reported having friends with divergent political opinions. This is in line with the notion that high levels of \textit{sociality} diversifies political discourse on social media platforms \cite{Brundidge}. 

Suspecting that social media platforms may facilitate the spiral-of-silence phenomenon on political issues, authors in \cite{Gearhart} revealed that "encountering agreeable political content predicts speaking out, while encountering disagreeable postings stifles opinion expression." Authors in \cite{Kwon} found that the fear of isolation from offline contacts increases the willingness to self-censor when it comes to posting political comments on Facebook. A recent study further confirmed the opinion congruence-based mechanism argued by the spiral-of-silence theory when expressing political opinions on Facebook \cite{Liu}. Authors in \cite{Fox} confirmed these findings when it comes to commenting on police discrimination on Facebook. Most recently, authors in \cite{Kushin} assessed the spiral-of-silence in the context of the 2016 US presidential election. Their analysis suggests that the more people perceived a public opinion support for Hillary Clinton, the less likely were to share a divergent pinion. This same phenomenon was particularly present for Donald Trump on Facebook. Because Donald Trump was highly unfavorable among Facebook users therefore inducing a spiral-of-silence among those who might have supported him in reality. These studies suggest the spiral-of-silence occurs when Facebook users discuss polarizing political issues. 

\section{Socially Engineering a Spiral-of-Silence on Facebook}
\subsection {Overview}
The spiral-of-silence theory posits that humans fear isolation, which motivates us to observe our social environment and mass media to determine opinion climate on issues with a strong moral component. The results of this observation influences our opinion expression in public, both interpersonally and on social media. The studies testing the the spiral-of-silence tenability assume that the public opinion climate is assessed from legitimate sources of information. What if this assumption is violated? The Cambridge Analytica incident provides reasonable grounds for us to believe that a malicious actor might resort to manipulating how one derives the public opinion climate about a polarizing political issue, for example, a presidential election or a referendum to leave the EU.

One option is by \textit{trolling}, a tactic where a malicious actor affects the public opinion climate by posting provoking and inflammatory messages and/or comments. Although effective in the past, social media platforms nowadays are taking active measures to curb trolling and remove suspicious user accounts and content. Another option for a malicious actor is to use MIM to covertly manipulate the public opinion climate for targeted set of social media users. Instead of infusing inflammatory content, the idea is to make a authentic posts and comments look "polarized." The MIM browser extension described above can be used for this purpose and alter valenced words in the comments section before it is presented in the targeted user's browser. The goal is to "socially engineer" the spiral-of-silence effect. In other words, the malware either induces or eliminates the fear of isolation and with that makes a target user more or less willing to self-censor their opinion. This motivated us to investigate whether an MIM attack will affect one's perception on the public opinion climate. 

\subsection {Research Questions and Hypotheses}
The study utilized a social media post about freedom of political expression on college campuses. This topic was chosen following President Donald Trump's executive order to protect freedom of speech on college campuses \cite{Svrluga}. Expressing political opinions on college campuses is a polarizing issue and has generated substantial media coverage and induced heated discussion both in-person and online \cite{Beauchamp}, \cite{Peters}. We opted out for this polarizing topic in the Facebook post to eliminate any a priori bias about a trending topic that a participant might have seen before. Another objective was to capture the initial reaction of the participants to a "new" post that was based on real, authentic events. The study also focused on one Facebook post with a limited number of comments instead of multiple posts to mimic a realistic setting where users qucikly skim a piece of online text, e.g. a "new" Facebook post ~\cite{Duggan}. 

The \textit{original scenario}, shown in Figure 3, included an authentic Facebook post about a report on political bullying at a higher education institution followed by authentic \textit{conservative-leaning} comments. The comments were from users with generic aliases and removed profile pictures to eliminate any potential bias on the grounds of popular trolling accounts. We used the malware to manipulate the comments and make them appear \textit{liberal-leaning} in the \textit{MIM scenario} shown in Figure 4. The malware replaced the words "liberal" with "conservative," "far-left" with "far-right," "over-parented" with "under-parented," "more" with "less," "far-left" with "far-right", and "Trump" with "Alexandria Ocasio-Cortez" (we took a surface polar opposite approach following the reports of both President Trump and congresswoman Ocasio-Cortez blocking users from their social media accounts \cite{Mays}, \cite{Savage}). 

Participants were randomly assigned in a control group (original scenario) and treatment group (MIM scenario). The collected data were used to investigate the possibility to "socially engineer" the spiral-of-silence effect for Facebook users. Because the four response strategies proposed in \cite{Hayes} indicate the predisposing spiral-of-silence conditions on social media, we used them as a primary dependent variable to explore whether the malware, by inducing misperceptions, can covertly nudge people to choose a particular one: \\

\textbf{Research Question 1}: \textit{How the manipulated Facebook post on the freedom of political expression on college campuses influences the utilization of different response strategies as predisposing spiral-of-silence conditions?} \\

To test the existence of the spiral-of-silence effect on Facebook, based on the predisposing conditions and using the well-established \textit{willingness to self-censor} measure \cite{Gearhart}, \cite{Liu}, we proposed the following hypothesis: \\

\textbf{Hypothesis 1}: \textit{The willingness to self-censor will be negatively related to publicly expressing an opinion in both the original and the MIM opinion climates (likelihood of commenting on the Facebook post).} \\

Because of the nature of the topic, there is a reason to suspect that the frequency one follows political news and uses social media may play an important role when deciding whether to comment or not. Therefore, we added the following hypotheses to our tests: \\

\textbf{Hypothesis 2a}: \textit{The frequency of following political news will be a strong predictor of the utilization of different opinion expression strategies on Facebook in both the original and the MIM scenarios.} \\

\textbf{Hypothesis 2b}: \textit{The frequency of use of social media will be a strong predictor of the utilization of different opinion expression strategies on Facebook in both the original and the MIM scenarios.} \\

Previous research on spiral-of-silence in social media explored the effects of opinion strength such as \textit{attitude certainty} (i.e. the degree to which one feels about their own opinion is correct \cite{Matthes1}) and \textit{perceived issue importance} (i.e. how important is the freedom of speech on college campuses to the general public \cite{Moy}), we asked: \\

\textbf{Research Question 2}: \textit{How will attitude certainty influence the utilization of different response strategies on Facebook when discussing the freedom of political expression on college campuses?} \\

\textbf{Research Question 3}: \textit{How will the perception of issue importance influence the utilization of different response strategies on Facebook when discussing the freedom of political expression on college campuses?} \\

There is also a reason to suspect that the perceived opinion climate of one's friends and family and of the nation may influence the willingness to express opinions, as found in some instances \cite{Matthes1}. These hypotheses tested these claims: \\

\textbf{Hypothesis 3a}: \textit{The perceived opinion climate among friends and family will be strong predictors of the utilization of different opinion expression strategies on Facebook when discussing the freedom of political expression on college campuses.}\\

\textbf{Hypothesis 3b}: \textit{The perceived opinion climate of the nation will be strong predictors of the utilization of different opinion expression strategies on Facebook when discussing the freedom of political expression on college campuses.}

\begin{figure}[h]
\centering
  \includegraphics[width=0.8\columnwidth]{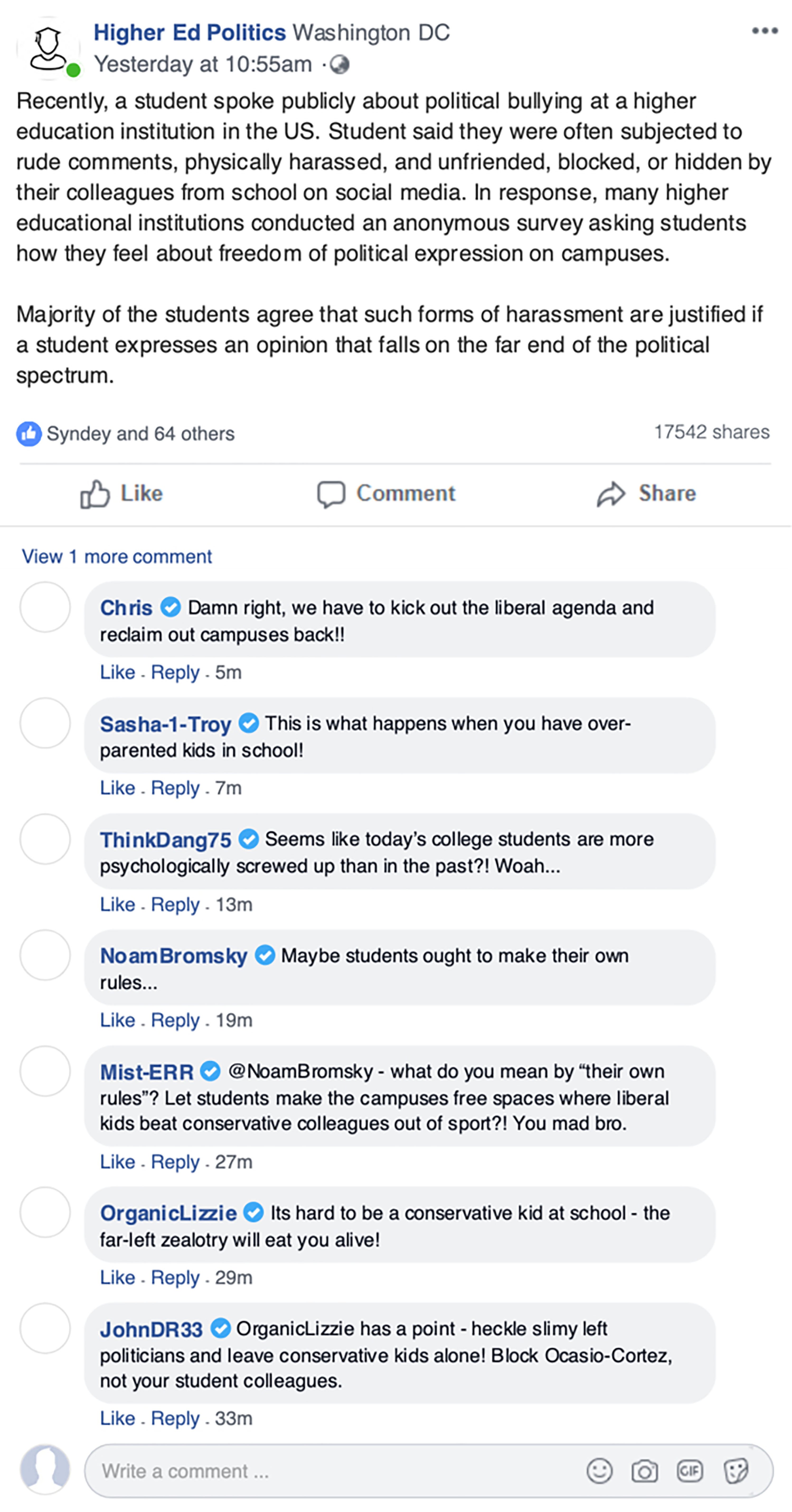}
  \caption{The original Facebook post and comments.}
  \label{Fig4}
\end{figure}
\newpage
\begin{figure}[h]
\centering
  \includegraphics[width=0.8\columnwidth]{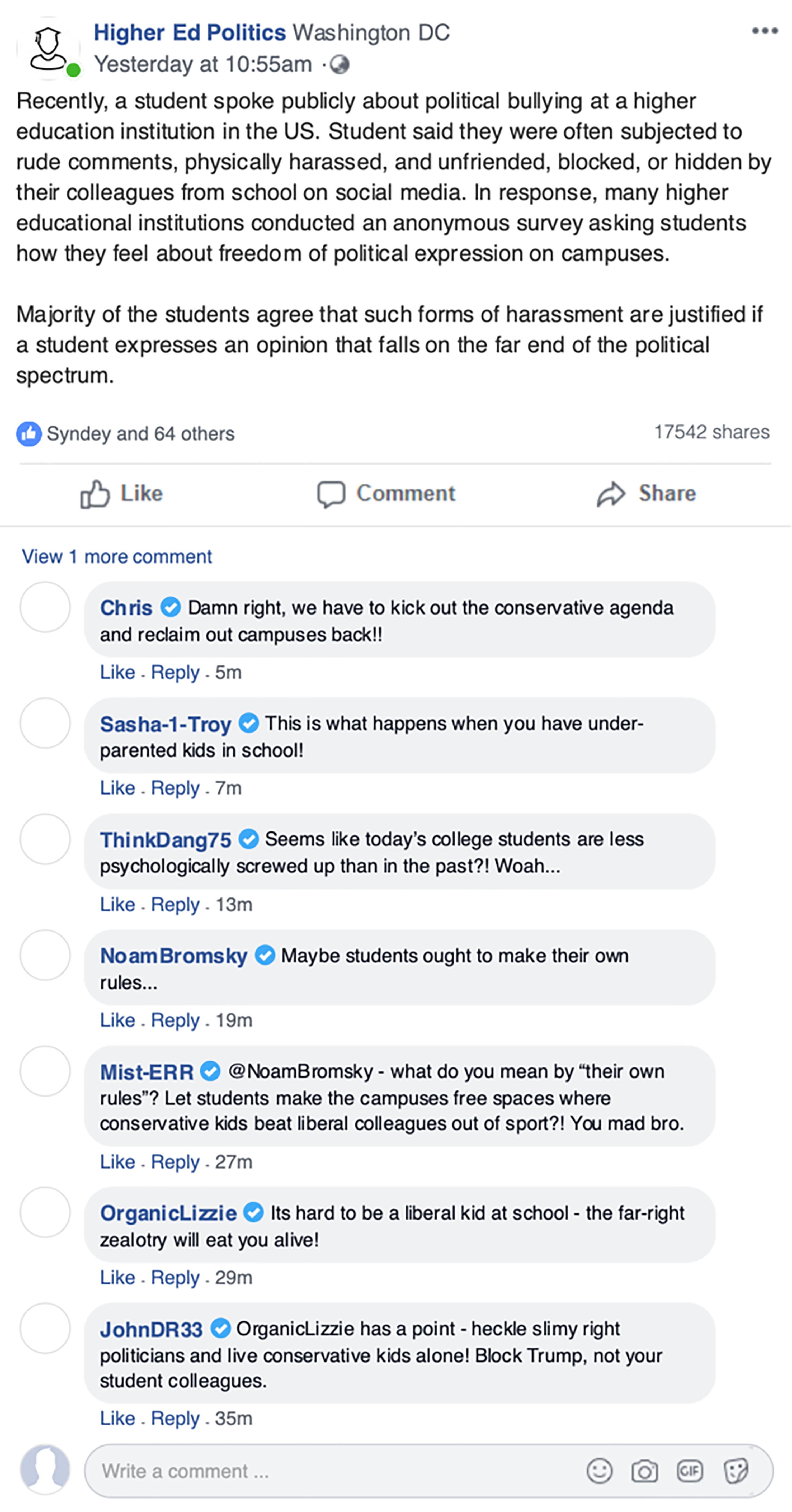}
  \caption{The MIM Facebook post and comments.}
  \label{Fig5}
\end{figure}

\section{Results}
Following an IRB approval, data were obtained through an online survey (N = 311), fielded via Prolific, a crowd-sourced participants pool \cite{Prolific}. Due to the choice of the topic of this study we recruited participants in the age bracket between 18 and 34 that are either college students or have recently earned a  bachelor's degree. The spiral-of-silence theory assumes that the topic is of personal relevance for an individual to engage with it, therefore the selection criteria required participants to be mainly college-age \cite{Newman}. Participants consisted of 55\% cis-female (N = 171), 42.1\% cis-male (N = 131), 0.3\% transgender female (N = 1), 1.3\% transgender male (N = 4), gender variant/non-conforming 1.0\% (N = 3), and 0.3\% preferring not to answer (N = 1). Participants were randomly assigned to either the original or the MIM scenario and completed a questionnaire. The questionnaire asked the participants about their response to the Facebook post and comments, about social media use, following of political news, opinions, attitudes, and the issue importance. Upon completion, participants were debriefed and rewarded a small monetary prize.

\subsection{Predisposing Spiral-of-Silence Conditions on Facebook} 
Research Question 1 explored how a manipulated Facebook post on the topic of freedom of political expression on college campuses influences the choice of a response strategy. As shown in Table 1 and 2, participants in the MIM scenario are more likely to comment on the Facebook scenario and post ($p = 0.031$) and more likely to tell someone else about it offline ($p = 0.038$) compared to the participant in the original scenario. The calculated effect size is  $<$ 0.3 (small). Participants didn't show any difference on the other two response strategies between the scenarios. As suspected, the results demonstrate that the malware is capable of engineering the spiral-of-silence effect on Facebook. This is an important finding that confirms the existence of the spiral-of-silence effect on a polarizing political issue on Facebook for the younger-leaning participant sample in our study. By inducing a misperception that the opinion climate is liberal-leaning, the malware  eliminated the fear of isolation from the original, conservative-learning scenario, and encouraged the participants to express their opinion, both online and offline. 

\begin{table}[htbp]
\renewcommand{\arraystretch}{1.5}
\caption{Matt-Whitney U Test for the Two Scenarios as a Grouping Variable.}
\label{table_example}
\centering
\begin{tabular}{|c|c|c|c|c|}
\hline
& \textbf{Comment} & \textbf{Read, not Comment} & \textbf{Ignore} & \textbf{Tell Offline}\\
\hline
$U$  & 10501 & 11892 & 10474 & 10914\\
\hline
$Z$  & 2.155  & .259 & 1.411 & 2.072\\
\hline
Sig & .031* & .796  & .158 & .038* \\
\hline
\multicolumn{5}{|l|}{*p $<$ .05, **p $<$ .01} \\
\hline
\end{tabular}
\end{table}

\begin{table}[h]
\renewcommand{\arraystretch}{1.5}
\caption{Descriptive Statistics For the Signficant Responses Strategies.}
\label{table_example}
\centering
\begin{tabular}{|c|c|c|c|c|}
\hline
& \multicolumn{2}{|c|}{\textbf{Comment}} & \multicolumn{2}{|c|}{\textbf{Tell Offline}}\\
\hline
 Scenario & \textbf{MIM} & \textbf{Original} & \textbf{MIM} & \textbf{Original} \\
\hline
Mean & 2.42  & 2.08 & 3.88 & 3.95 \\
\hline
Median & 2 & 1  & 4.0 & 4.5 \\
\hline
STD & 1.732  & 1.64 & 1.810 & 1.857 \\
\hline
\end{tabular}
\end{table}

This is an expected outcome in the context of political discourse, i.e. the malware covertly created the necessary conditions that allowed the participants in the MIM scenario to succumb to the characteristic 'echo chamber' effect \cite{Gillani}. Seeing a favorite narrative in the MIM scenario reinforced the confirmation bias that helped to account for participants’ decisions about whether to spread content both online and offline, as the formative action that leads to towards preferential interaction based on confirming claims in the Facebook comments section of the post \cite{Quattrociocchi}. The preferential interactions, based on the malware-induced misperception, initiate a feedback loop that continuously amplifies ideologically orthodox comments and posts and drowns out any opposing views, ultimately resulting into the spiral-of-silence effect \cite{Noelle-Neumann}. 

\subsection{Socially Engineered Spiral-of-Silence on Facebook}
Hypothesis 1 claimed that the willingness to self-censor, as a composite measure, will be negatively related to the likelihood to comment on the Facebook post in both scenarios. Based on Table 3, the more participants were willing to self-censor, the less likely they were to publicly comment on the Facebook post (original condition $\beta= -.228$, $p < .01$; MIM condition $\beta = -.334$, $p < .01$), confirming the prediction in Hypothesis 1. These results demonstrate the existence of the spiral-of-silence effect on Facebook on the particular issue investigated in our study in both the original and MIM scenario, proving the capability of the malware to induce misperception of the public opinion climate \textit{without} raising suspicion. This is a very important finding that demonstrates the capability of the malware to socially engineer the spiral-of-silence effect on social media, Facebook in particular. In addition, the results also confirm the previous evidence that individuals with high levels of willingness to self-censor use Facebook as a forum to monitor public opinion on important social and political issues when expressing their opinion offline ~\cite{Gearhart}, \cite{Kwon}, \cite{Liu}.

\begin{table}[h]
\renewcommand{\arraystretch}{1.7}
\caption{Hierarchical Regression Predicting the Likelihood of Commenting on the Facebook Post.}
\label{table_example}
\centering
\begin{tabular}{|l|c|c|c|c|}
\hline
 & \textbf{Original} & \pbox{20cm}{\textbf{Std.}} & \textbf{MIM} & \pbox{20cm}{\textbf{Std.}}\\
\hline
\multicolumn{5}{|l|}{\textbf{Demographics}} \\
\hline
Age & .026 & .001 & .349 & .036\\
\hline
Gender & .208 & .088 & -.167 & -.092 \\
\hline
Incr. $R^2 (\%)$ & \multicolumn{2}{|l|}{3.5} & \multicolumn{2}{l|}{4.7}\\
\hline
\multicolumn{5}{|l|}{\textbf{Social Media and Politics}} \\
\hline
Social Media Use & .194 & .097 & -.042 & -.016\\
\hline
Following Politics & -2.45* & -.169* & -.276* & -.190* \\
\hline
Incr. $R^2 (\%)$ & \multicolumn{2}{|l|}{5.2**} & \multicolumn{2}{l|}{5.9*}\\
\hline
\multicolumn{5}{|l|}{\textbf{Focal Variables}} \\
\hline
\pbox{5cm}{Willingness to \\ \vphantom{g} self-censor} & -0.512** & -.228** & -.831** & -.334**\\
\hline
 \pbox{5cm}{Attitude certainty} & .036 & .034 & .001 & .001\\
\hline
 \pbox{5cm}{Issue importance} & -.241 & -.131 & .145 & .072\\
\hline
 \pbox{5cm}{Congruence \\ \vphantom{g} friends \& family} & .002 & .034 & .002 & .033 \\
\hline
 \pbox{5cm}{Congruence \\ \vphantom{g} nation} & .000 & .004 & .004 & .048\\
\hline
Incr. $R^2 (\%)$ & \multicolumn{2}{|l|}{6.8*} & \multicolumn{2}{l|}{11.3*}\\
\hline
Total $R(\%)$ & \multicolumn{2}{|l|}{15**} & \multicolumn{2}{l|}{21.9**}\\
\hline
\multicolumn{5}{|l|}{*p $<$ .05, **p $<$ .01} \\
\hline
\end{tabular}
\end{table}

Hypothesis 2a claimed that the frequency with which one follows political news will be strong predictors of the utilization of different opinion expression strategies. The frequency of following political news was measured by asking, "How closely do you follow political news" (1 = Never  to 5 = Always; $M = 2.85$, $SD =1.155$). Based on Tables 3-6, the more frequently one follows political news: 

\begin{itemize}
	\item the less likely is to comment on the Facebook post in both scenarios (original scenario $\beta= -.169$ $p < .05$; MIM scenario $\beta = -.190$, $p < .05$)
	\item the less likely to read but not comment on the Facebook post in both scenarios (original scenario $\beta= -.277$ $p < .01$; MIM scenario $\beta= -.194$, $p < .01$)
	\item the more likely to ignore the Facebook post in both scenarios (original scenario $\beta= -.390$ $p < .01$; MIM scenario $\beta= - .353$, $p < .01$)
	\item the more likely to tell someone about the Facebook post in the original scenario ($\beta= .142$ $p < .01$)
\end{itemize}

These results shed further light into the capabilities of the malware and the possibility of profiling future MIM targets. The stronger negative relationship between the frequency of following political news and and the first response strategy (Table 3) in the MIM scenario compared to the weaker relationships for the other response strategies (Table 4-6) indicates that the primary targets for the MIM attacks should be the individuals that are interested in the daily politics but remain largely "undecided." This is a well known fact that is used in political campaigning well before social media became a factor in inducing voter bias \cite{Benoit}. The results also indicate that the MIM attack is not simply an alternative to trolling, but it is a much more powerful tool for influencing outcomes. The social media trolls usually target individuals that follow political news with high frequency; MIM on the other hand, allows for targeting the individuals without a particular pattern of daily check-ups for the public opinion climate. 

Hypothesis 2b claimed that the frequency with which one uses social media will be strong predictors of the utilization of different opinion expression strategies. The frequency of using social media was measured by asking, "How often do you use social media" (1 = Never  to 5 = Several Times a Day; $M = 4.53$, $SD =0.74$). Based on Tables 3-6, the more frequently one uses social media:

\begin{itemize}
	\item the more likely to read but not comment on the Facebook post only in the original scenario ($\beta= .352$ $p < .01$)
	\item the less likely to ignore the Facebook post only in the original scenario ($\beta= .142$ $p < .01$)
\end{itemize}

The frequency of social media use proved not to be a decisive predictor in speaking up or silencing in the MIM scenario. The same can be concluded for the original scenario given that significance is achieved only for the response strategies that ignore or simply read the post and comments. Seeing this result from a profile perceptive, as discussed before, the MIM attackers need not to worry about \textit{how} frequently one uses social media, but for \textit{what} purpose. This uncovers another utility of the MIM attack - it can be used, in a same fashion as trolling, if the attackers choose to alter the factual integrity of the Facebook content and post and make it look more provoking or sound inflammatory to the target users. 

\begin{table}[h]
\renewcommand{\arraystretch}{1.7}
\caption{Hierarchical Regression Predicting the Likelihood of Reading but not Commenting the Facebook Post.}
\label{table_example}
\centering
\begin{tabular}{|l|c|c|c|c|}
\hline
 & \textbf{Original} & \pbox{5cm}{\textbf{Std.}} & \textbf{MIM} & \pbox{5cm}{\textbf{Std.}}\\
\hline
\multicolumn{5}{|l|}{\textbf{Demographics}} \\
\hline
Age & 2.137 & .104 & 1.99 & .210 \\
\hline
Gender & -.299 & -.122 & .095 & .053 \\
\hline
Incr. $R^2 (\%)$ & \multicolumn{2}{|l|}{0.1} & \multicolumn{2}{l|}{0.7}\\
\hline
\multicolumn{5}{|l|}{\textbf{Social Media and Politics}} \\
\hline
Social Media Use & .677** & .352** & .291 & .11 \\
\hline
Following Politics & -.388** & -.277** & -.286* & -.194*\\
\hline
Incr. $R^2 (\%)$ & \multicolumn{2}{|l|}{21.0**} & \multicolumn{2}{l|}{5.9*}\\
\hline
\multicolumn{5}{|l|}{\textbf{Focal Variables}} \\
\hline
\pbox{5cm}{Willingness to \\ \vphantom{g} self-censor} & -012 & -005 & .178 & .070\\
\hline
 \pbox{5cm}{Attitude certainty} & .134 & .128 & .018 & .015 \\
\hline
 \pbox{5cm}{Issue importance} & -.108 & -.061 & .239 & .117 \\
\hline
 \pbox{5cm}{Congruence \\ \vphantom{g} friends \& family} & .010 & .162 & .000 & -.004 \\
\hline
 \pbox{5cm}{Congruence \\ \vphantom{g} nation} &  -.007 & -098 &  -.001 & -.008 \\
\hline
Incr. $R^2 (\%)$ & \multicolumn{2}{|l|}{2.5} & \multicolumn{2}{l|}{1.9}\\
\hline
Total $R(\%)$ & \multicolumn{2}{|l|}{23.6} & \multicolumn{2}{l|}{8.5}\\
\hline
\multicolumn{5}{|l|}{*p $<$ .05, **p $<$ .01} \\
\hline
\end{tabular}
\end{table}

\begin{table}[h]
\renewcommand{\arraystretch}{1.7}
\caption{Hierarchical Regression Predicting the Likelihood of Ignoring the Facebook Post.}
\label{table_example}
\centering
\begin{tabular}{|l|c|c|c|c|}
\hline
 & \textbf{Original} & \pbox{5cm}{\textbf{Std.}} & \textbf{MIM} & \pbox{5cm}{\textbf{Std.}}\\
\hline
\multicolumn{5}{|l|}{\textbf{Demographics}} \\
\hline
Age & -1.05 & -.005 & .034 &.003 \\
\hline
Gender & -.025 & -.009 & .242 & .127 \\
\hline
Incr. $R^2 (\%)$ & \multicolumn{2}{|l|}{0.3} & \multicolumn{2}{l|}{0.9}\\
\hline
\multicolumn{5}{|l|}{\textbf{Social Media and Politics}} \\
\hline
Social Media Use & -.474** & -.214** &  -.114 & -0.41\\
\hline
Following Politics & .663** & .390** & .546** & .353**\\
\hline
Incr. $R^2 (\%)$ & \multicolumn{2}{|l|}{19.2**} & \multicolumn{2}{l|}{12.4**}\\
\hline
\multicolumn{5}{|l|}{\textbf{Focal Variables}} \\
\hline
\pbox{5cm}{Willingness to \\ \vphantom{g} self-censor} & -.012 & -.005 & .330 & .125\\
\hline
 \pbox{5cm}{Attitude certainty} & .194 & .161 & -.068 & -.056\\
\hline
 \pbox{5cm}{Issue importance} & -.122 & -.059 &  .208  & .097 \\
\hline
 \pbox{5cm}{Congruence \\ \vphantom{g} friends \& family} & -.003 &  -.041 & .002 & .037 \\
\hline
 \pbox{5cm}{Congruence \\ \vphantom{g} nation} & -.002 & -.024 & -.004 & -.043\\
\hline
Incr. $R^2 (\%)$ & \multicolumn{2}{|l|}{2.4} & \multicolumn{2}{l|}{2.8}\\
\hline
Total $R(\%)$ & \multicolumn{2}{|l|}{29.1} & \multicolumn{2}{l|}{16.1}\\
\hline
\multicolumn{5}{|l|}{*p $<$ .05, **p $<$ .01} \\
\hline
\end{tabular}
\end{table}

The versatility of the MIM attack is further corroborated with the results of the tests of Hypothesis 3a and 3b show in in Table 3-6. The claims that the perceived opinion climate among friends and family and among the nation, respectively, will be strong predictors of the utilization of different opinion expression strategies were unsupported in our particular case. Similarly, we haven't found evidence that the attitude certainty (Research Question 2) and the perceived issues importance (Research Question 3) influence the utilization of the response strategies. Seeing this result from a profile perceptive, the MIM attackers need not to worry about what the target users talks with their friends and family or whether the user believes the issue is important to the general public. Looking back to the findings from Research Question 1, these results confirm that the MIM attack is only concerned about the search for confirming claims in the Facebook comments section of the post (the 'echo chamber' effect). This means that it is sufficient for the malware to induce misperceptions about the "majority" opinion climate without considering any other factors in order to socially engineer the spiral-of-silence effect on social media. The overall findings make a compelling case for a resourceful actor, interested in alternative to trolling, to invest into developing and disseminating a misperception-inducing malware.


\begin{table}[h]
\renewcommand{\arraystretch}{1.7}
\caption{Hierarchical Regression Predicting the Likelihood of Telling Someone Else Offline about the Facebook Post.}
\label{table_example}
\centering
\begin{tabular}{|l|c|c|c|c|}
\hline
 & \textbf{Original} & \pbox{5cm}{\textbf{Std.}} & \textbf{MIM} & \pbox{5cm}{\textbf{Std.}}\\
\hline
\multicolumn{5}{|l|}{\textbf{Demographics}} \\
\hline
Age & 2.607 & .112 & -.446 & -.043 \\
\hline
Gender & -.092 & -0.33 & -.161 & -.043 \\
\hline
Incr. $R^2 (\%)$ & \multicolumn{2}{|l|}{1.9} & \multicolumn{2}{l|}{0.5}\\
\hline
\multicolumn{5}{|l|}{\textbf{Social Media and Politics}} \\
\hline
Social Media Use & .254 & .112 & .026 & .009 \\
\hline
Following Politics & -.412** & .142** & -.273 & -.171\\
\hline
Incr. $R^2 (\%)$ & \multicolumn{2}{|l|}{9.2**} & \multicolumn{2}{l|}{3.3}\\
\hline
\multicolumn{5}{|l|}{\textbf{Focal Variables}} \\
\hline
\pbox{5cm}{Willingness to \\ \vphantom{g} self-censor} & .279 & .110 & .126 & .046\\
\hline
 \pbox{5cm}{Attitude certainty} & -.066 & -.054  & .113 &  .090\\
\hline
 \pbox{5cm}{Issue importance} & .230 & .110 & .107 & .048\\
\hline
 \pbox{5cm}{Congruence \\ \vphantom{g} friends \& family} &  -.004 &  -.062  & -.006 & -.092\\
\hline
 \pbox{5cm}{Congruence \\ \vphantom{g} nation} &  .013 & .167 & -0.004 & -.040\\
 \hline
Incr. $R^2 (\%)$ & \multicolumn{2}{|l|}{3.4} & \multicolumn{2}{l|}{2.1}\\
\hline
Total $R(\%)$ & \multicolumn{2}{|l|}{14.4} & \multicolumn{2}{l|}{5.9}\\
\hline
\multicolumn{5}{|l|}{*p $<$ .05, **p $<$ .01} \\
\hline
\end{tabular}
\end{table}

\section{Discussion}
This study, to our knowledge, is the first one to test the possibility of socially engineering or disrupting the spiral-of-silence on social media by employing a malware-induced misperception in a polarized discourse on Facebook. Previous studies exploring the spiral-of-silence effect assumed that individuals' perception of the public opinion is based on media information from authentic and credible sources. In our study, we used a malware to induce misperception by manipulating the linguistic formatting of authentic social media information, a post and comments discussing a polarizing political issue. The Cambridge Analytica scandal and the alleged Russian meddling with the 2016 elections provided an additional impetus for the test in order to scope the potential strategies for political influence before the election year 2020. 

Our initial tests demonstrate that a malware could successfully induce a misperception about the public opinion climate gauged from the people's interaction on social media. In our study, this malware covertly manipulated words in the conservative-leaning comments section of a Facebook post to make them appear liberal-leaning, and with that, created a the perception for our liberal-leaning sample that the opinion climate is preferential to them. Seeing a favorite narrative, participants took a formative action towards sharing their opinion both online and offline. Our further analysis demonstrated that the misperception induced by the malware was sufficient to socially engineer the spiral-of-silence effect. The preferential interactions of most participants were to talk about the Facebook post offline instead of online, which initiate a feedback loop that continuously amplifies  ideologically liberal comments and posts and drowns out any opposing views, ultimately resulting into the spiral-of-silence effect. 

In other words, the malware preliminary disrupted the predisposing spiral-of-silence conditions to nudge participants to succumb to the 'echo chamber' effect, and with that, avoid to share their opinion publicly online. The findings of the study supports the claim that "engaging in opinion expression to someone offline removes the inherent risks associated with expressing opinions in a public online forum composed of people one knows in real life" \cite{Gearhart} ,\cite{Liu}. This is an important notion from a political influence perspective because it confirms the findings that "social media users, despite being reluctant to publicly comment on the post, are actively engaged in this environment through observation" \cite{Gearhart}, \cite{Kwon}. 

The MIM attack works in a highly targeted fashion and has a reduced reach compared to other forms of online influence like trolling. Target profiling, then, is more important to a MIM attacker and we also conducted an analysis to see the profile of targets that will mostly fall victim to a MIM attack. Our analysis suggests that the most likely victims to the MIM attacks are the people who follow political news, but remain generally undecided on most polarizing issues on social media. Demographic aspects like age, gender, and social media use have shown in our analysis to be irrelevant factors. It also doesn't matter for MIM attackers if a target user's opinion is congruent with the public opinion. Victims to the MIM attack can also be anyone regardless of their attitude certainty or perceived issue importance on the particular polarizing discourse and issue of freedom of speech on college campuses.

\subsection {Implications}  
The malware, as demonstrated, has the potential to "nudge" a target user to focus on the opinion climate rather than assessing whether a Facebook post and comments are intended as trolling or rumors. The MIM attack vector, in other words, is not aimed at the social media platform but rather at a user or group of users of interest. This eliminates the constraint that the platform administrators will remove suspicious content and places the burden of defense on the user side. The MIM can be used to "socially engineer" a targeted user to break out from the spiral-of-silence and express their opinion on a topic that, under normal conditions, they would choose not to say offline. The alternative outcome is also possible: silencing users on topics upon which they would usually choose to express on social media. For posts with public comments, the MIM attack can work with minimal to no adaptation for more than one political topic (e.g. foreign policy, immigration, tariffs, and reproductive health). This allows the malicious actors to dynamically re-purpose the attack depending on the trending political discourse on the social media platforms. 

The ethical implications of our MIM study are the same as those related to publishing any vulnerability: the value of publicly sharing a proof-of-concept social engineering attack with knowledgeable researchers outweighs the opportunity that potential attackers may benefit from the publication. If this paper introduces a viable attack in the social media ecosystem--which it might will--due to its simplistic nature, we believe that this might be merely a confirmation of similar attacks, independently developed and deployed by well-resourced adversaries or nation-state groups. The study itself tests the plausibility of a locally developed MIM browser extension (not publicly available on the Chrome store). In the context of a real-life MIM attack, a responsible disclosure would entail contacting Google, the developers of Chrome, and working with them through the details of the malware extension.

\subsection {Limitations} 
Though the results of this study suggest that the MIM attack is capable of socially engineering or disrupting the spiral-of-silence within social media on a polarizing political issue, caution is warranted when interpreting them. The use of controlled Facebook post allowed us to capture the first impressions of the participants, but this choice at the same time limits the generalization of the findings in regards to the real opinion expression behavior. The polarizing topic chosen in this study might have been of variable degree of interest to individual participants, which also affects their decision to express their opinion. We tried to control for this by selecting a younger, college-age population assuming that the issue of freedom of political expression is highly relevant for them and they can identify with it. This on the other side, limits the generalization of the findings about an older population that has a more distant outlook on this issue considering other factors such the general political climate or their attitude certainty. Same holds for the self-reported frequency of following political news, which may be influenced by the type of news, outlets, topics, and interfaces. We anonymized the comments in both scenarios and didn't explicitly ask whether participants will express their opinion if anonymity is granted. Anonymity is an integral part of the social media ecosystem and further research should test the MIM potential of socially engineering a spiral-of-silence process under conditions of anonymity. 

Our results are also limited particular choice of web browser as an interface and a particular social media site - Facebook. The malware was tested in its extension variant but there are many people that access social media through smartphone applications or multiple interfaces in the same time. There is a possibility that the same results might not be obtained because smartphone applications provide a different set of interaction affordances that limit the cues one uses to access the opinion climate. Similarly, using multiple interfaces contributes to repetitive exposure to the same information which can lead to changes in perceptions about the issue importance and one's attitude certainty. This, in turn, can make people more or less compelled to comment on a polarizing issue regardless of their political ideology or gender identity. The particular choice of social media site also limits the generalization because other social media platforms have different affordances that influence one's opinion formation and decision to speak out. For example, Twitter has limited text input, Instagram is heavy on non-textual content (e.g. images, videos, gifs), while Reddit has ‘SubReddits,’ ‘up’ or ‘down-voting’, and the act of giving ‘gold’. Because these affordances shape norms of what people share and expect to see being shared, different platforms could have a variable degree of conductivity to a socially engineered spiral-of-silence effect.

The sample in the study was liberal-leaning and the findings might be different for a representative sample. We didn't control for any other dimensions of one's political identity, which certainly factor in one's willingness to self-censor. For example, individuals' partisanship, structure, culture, and historical experience of society often shape the preconceptions of a polarizing issue at stake, even in circumstances where the people put a premium on purportedly independent and objective public opinion assessment \cite{Simpson}. On this token, MIM is a novel attack and users are unaware of its existence to be able to detect it in the first place, regardless of any prior phishing training or negative experience with trolling and propaganda on social media. The outcomes of the study may be different if user awareness about this attack is raised, as it is usually the case with social engineering attacks. Although we demonstrated the potential of the MIM attack, it might be hard to scale it up quickly to a large social media population like the trolling, rumor, or disinformation campaigns do, but that is what makes the MIM attack compelling to a malicious actor.

\subsection {MIM Defenses and Prevention}
The study introduces a plausible social engineering vector against individuals and groups that has yet to emerge in the wild, but has analogs in other deception and information warfare contexts \cite{Cronin}. The threat of MIM is an inherent risk of computer-mediated communication, particularly as artificial intelligence and machine learning enable software to parse and edit text toward particular opinion climate, emotional tone, or adversarial perspective. The first line of defense would require elimination of any suspicious extensions in the Chrome store that require permissions to control how HTML text is presented to a user. An example defense, along the lines of malicious software detection, would be using trusted browsers to detect JavaScript executions that are rearranging words and sentences in the textual portion of an HTML document \cite{Kohlbrenner}. Another example is Chrome's Manifest v3 API, which is designed to eliminate extensions exhibiting suspicious behaviour in content manipulation \cite{Google}. Content-level signing might not help in these regards because the MIM manipulation happens after the content integrity check in the sequence of HTML reception and display.

One thing to have in mind is the possibility of the sneaking the malware extension on the Chrome store as an "accessibility (a11y) extension" by claiming that the rewording is done to create an assistive natural language software that, for example, helps non-native English speakers \cite{Jang}. It might be harder to bar an extension from the Chrome store on these grounds, therefore, the certification process must request all the use cases for these word manipulations upfront to ensure no misperception-inducing logic is hidden in the inner workings of the assistive extension. Even with these cautions, a malicious actor may find a way to deploy the malware on a target's browser (for example, an insider threat).

As with any social engineering tactic, awareness of the potential attack is an advantage to the defender and a second line of defense. Given that the attack takes place on the target's browser and not the social media platforms, this might be the only available option for individuals at this point. A practical training session for detecting MIM attacks revolves around the idea of crossing the \textit{deception judgment threshold}, as argued by the Truth-Default theory and scrutinizing the Facebook post and comments \cite{Levine}. The traditional social engineering training is focused on quick visual assessments for the most reliable indicators like URLs, grammar, padlocks for https, links, and attachments. This is already in place for the MIM attack. The focus for the MIM training is thus on the analysis of the social media posts and comments in the broader context of the issue at stake (in our case, freedom of speech on college campuses). The deception judgment can be calibrated based on updated facts for both the perceived majority and minority opinions. Compared to the traditional social engineering victims, the MIM victims have the advantage of individually approaching each of the people that commented on the post and verifying their original opinion. Or, verifying the authenticity of the comments and the prevailing public opinion by checking other media sources reporting on a given polarizing issue. Certainly, this out-of-band verification might make the social media interaction cumbersome, but that is a very small cost to quickly cross the deception judgment threshold. We believe this is an empowering strategy, and suggest that any social engineering training has a section on MIM as a tactic for inducing misperception on social media. 

\subsection{Future Work}
For our next research steps we plan to replicate and extend the current study with other social media sites (e.g. Twitter, Reddit) to explore whether the affordances of a particular social media side affects the choice of a response strategy. Our plan is also to cover other controversial topics popular or social media, for example vaccination, conspiracy theories or global warming that to not necessarily divide the people on political ideology or gender identity lines. We will work on diversifying our future samples and control for age, level of education, or other demographic and cultural factors so as to get a more nuanced idea on how a spiral-of-silence effect, socially engineered or disrupted by a covert malware, might unfold in the future for a purpose of a covert, low-intensity political propaganda. Towards a more robust test of the malware, the future research will investigate whether a different packaging, e.g. a third-party smartphone social media application, could amplify or attenuate the misperception-inducing potential of the malware. Another line of research will continue to explore machine learning mechanism for automated decision making on what type of linguistic rearrangement is the best suited for a particular polarizing issue, target, or a social media platform. Our objective in future research is not to perpetuate any deviant cybersecurity behaviour, quite the contrary. We are strongly dedicated to investigating any facet of the MIM attack to be able to eradicate it with both technological and societal prevention mechanisms. 

\section{Conclusion}
In this work, we introduced the MIM attack as a means of covert opinion manipulation of political discourse on Facebook. We tested it with 311 participants and showed that the MIM attack has the potential to socially engineer the spiral-of-silence effect on social media. The results also show that the MIM attack has the potential to disrupt the spiral-of-silence by creating misperceptions about the public opinion climate and nudging people to succumb to the echo chamber effect. Our main contribution is the evidence that the spiral-of-silence effect can be induced on demand - only with a piece of seemingly benign JavaScript (or other software) code and without fabricating any social media content. We hope our results inform the security community about the implications of having an alternative social engineering vector for social media influence, at least in a micro-targeted variant. We are aware that malware and the attack have a long way to go before materialize into a sizable threat. Nevertheless, the early proof-of-concept demonstrated in this paper facilitates a critical, scientific outlook on the use covert malware in situations where social interaction is a decision making factor.

\bibliographystyle{IEEEtranS}

%

%
%

\bibliography{ATD_NDSS2020}






\end{document}